\documentclass[aps,showpacs,longbibliography,notitlepage,superscriptaddress,twocolumn,nofootinbib]{revtex4-1}
\pdfoutput=1
\usepackage{dsfont}
\usepackage{enumitem} 
\usepackage[utf8]{inputenc}
\usepackage[english]{babel}
\usepackage[T1]{fontenc}
\usepackage{amsmath}
\usepackage{xcolor}
\colorlet{myPurple}{blue!40!red}
\colorlet{myCyan}{cyan!60!gray}
\colorlet{myRed}{blue!55!gray}
\usepackage{tikz}
\usepackage{pgfplots}
\pgfplotsset{compat=1.14}
\usepackage[colorlinks=true,citecolor=myRed,urlcolor=myRed,linkcolor=myRed]{hyperref}
\usepackage{exscale}
\usepackage{fouriernc}
\usepackage{bbm}
\usepackage{graphicx}
\usepackage{amsmath}
\usepackage{latexsym}
\usepackage{amssymb}
\usepackage{mathtools}
\usepackage{amsfonts}
\usepackage{physics}
\usepackage{amssymb}
\usepackage{times}
\usepackage[T1]{fontenc}
\usepackage{amsthm}
\usepackage{enumerate}
\usepackage{csquotes}
\usepackage{bbold}
\usepackage{color}
\usepackage{nicefrac}
\usepackage{changes}
\usepackage{calrsfs}
\usepackage{soul}

\newcommand{\avg}[1]{\langle{#1}\rangle}

\usepackage{tcolorbox}

\usepackage{tikz}
\usepackage{lipsum}
\theoremstyle{plain}

\usepackage{graphicx}
\usepackage{bm}
\usepackage{dsfont}
\usepackage{tikz}
\usepackage[T1]{fontenc}
\usepackage{amsthm}
\usepackage{array}
\usepackage{amssymb}
\usepackage{amsfonts}
\usepackage{cancel}
\usepackage[toc,page]{appendix}
\usepackage{multirow}
\usepackage{color}
\usepackage{calrsfs}
\DeclareMathAlphabet{\mathpzc}{OT1}{pzc}{m}{it}
\usetikzlibrary{backgrounds,decorations.pathreplacing,calc}
\definecolor{mygray}{gray}{0.8}
\DeclareMathAlphabet{\mathcal}{OMS}{cmsy}{m}{n}

\makeatletter
\def\p@subsection{}
\makeatother

\begin{document}
\title{A perspective on few-copy entanglement detection in experiments}

\author{Valeria Saggio}
\thanks{valeria.saggio@univie.ac.at}
\affiliation{University of Vienna, Faculty of Physics, Vienna Center for Quantum Science and Technology (VCQ), Boltzmanngasse 5, 1090 Vienna, Austria}

\author{Philip Walther}
\affiliation{University of Vienna, Faculty of Physics, Vienna Center for Quantum Science and Technology (VCQ), Boltzmanngasse 5, 1090 Vienna, Austria}
\affiliation{Christian Doppler Laboratory for Photonic Quantum Computer, Faculty of Physics, University of Vienna, Vienna, Austria}

\begin{abstract}
Although the realization of useful quantum computers poses significant challenges, swift progress in emerging quantum technologies is making this goal realistically approachable. In this context, one of the essential resources is quantum entanglement, which allows for quantum computations outperforming their classical counterparts. However, the task of entanglement detection is not always straightforward for several reasons. One of the main challenges is that standardly-used methods rapidly become unfeasible when dealing with quantum states containing more than a few qubits. Typically, this is due to the fact that a vast amount of measurements is needed on many copies of the state. Generally, it is not unusual to deal with a very limited number of state copies in experimental settings - in fact, this may be the case for many large quantum systems. In this article, an overview is provided of a probabilistic approach that enables high-confidence genuine multipartite entanglement detection using an exceptionally low number of state copies. Additionally, a study is presented that shows that this protocol remains efficient also in the presence of noise, thus confirming the practicality of the method for near-term quantum devices and its suitability for complex experimental settings.
\end{abstract}

\maketitle

\section{Introduction}
The concept of quantum entanglement initially arose in the famous thought experiment of A. Einstein, B. Podolsky and N. Rosen as the cause of an apparent paradox explained with assuming the incompleteness of quantum mechanics~\cite{einstein1935can}. However, about 30 years later J. S. Bell provided a solution to the paradox deriving an inequality that, if violated, would prove against quantum mechanics being a local-realistic theory (like Einstein, Podolsky and Rosen assumed)~\cite{bell1964einstein}. The first experimental demonstrations of the violation of Bell's inequality~\cite{aspect1981experimental,aspect1982experimental,aspect1982experimental_}, realized by performing measurements on pairs of entangled particles, confirmed that no local theory could reproduce the predictions of quantum mechanics. These experiments were truly a breakthrough in the quantum field. They opened the way to a better understanding of the phenomenon of entanglement and led to analyses of its implications. Nowadays it is well established that entanglement plays a central role in efficient quantum computation~\cite{jozsa2003role}. Examples of quantum protocols requiring entanglement as an essential resource are many quantum communication and quantum information processing schemes~\cite{morales2015quantum,bennett1993teleporting,bouwmeester1997experimental,pan1998experimental} or algorithms based on the so-called \textit{measurement based quantum computing} (MBQC)~\cite{briegel2009measurement,walther2005experimental}, to name but a few. Given the key role that entanglement plays in quantum computation, it is evident that its characterization and quantification is equally important~\cite{friis2019entanglement}. In other words, as it is essential to verify that the entangled system used to perform a certain computational task operates in the anticipated fashion --- i.e. below some error threshold --- reliable verification of quantum entanglement becomes a crucial task. While it is relatively straightforward to reliably perform this task in small-scale quantum systems (many schemes have been developed that are easily applicable to quantum states with few qubits) it is also evident nowadays that this is not yet the case for systems containing a larger number of qubits. In fact, many entanglement detection methods are only practical at the very small scale of noisy intermediate-scale quantum (NISQ) devices (which contain tens of qubits) and are ill-suited for systems of a few hundred qubits, which we can expect in the next decade or so in a huge variety of quantum platforms, from photons to trapped ions and superconducting qubits~\cite{preskill2018quantum}. The reason as to why many methods are not efficiently applicable to large-scale NISQ and beyond NISQ devices lies in the amount of resources required. Ideally, the gold standard for every quantum system would be complete quantum state tomography, from which one can infer full information about the state (as this method fully reconstructs its density matrix)~\cite{james2005measurement}. Unfortunately, since the number of required measurement settings scales exponentially with the size of the system, i.e. number of qubits, this approach comes at a very high price. Although efforts were made to exploit certain state symmetries and thus achieve a polynomial scaling~\cite{toth2010permutationally} or to use compressed sensing schemes to reduce the high experimental demand~\cite{gross2010quantum}, this approach is still highly impractical for reconstructing large-scale states. Luckily, full information about the state is often not needed, and techniques that aim to probe only some specific features have been developed, generally leading to less costly experimental requirements. Examples of such tasks are the reconstruction of arbitrary elements of the quantum state density matrix~\cite{morris2019selective}, its projections onto a fixed set of observables~\cite{aaronson2019shadow}, its approximate classical description~\cite{huang2020predicting}, the estimation of how far an experimentally prepared state is from an ideal target state~\cite{flammia2011direct}, or the detection of entanglement~\cite{guhne2009entanglement}. As long as noise-resistant quantum computers are not fully developed, it is therefore essential to reliably validate any available quantum system. In the course of this article we will focus on the task of entanglement detection, providing an overview of the probabilistic few-copy approach presented in Ref.~\cite{saggio2019experimental} that enables entanglement detection with high confidence using an exceptionally low number of state copies. We will first present the theoretical background, and then provide an analysis of the tolerance to noise aimed at investigating how the amount of required resources varies in the presence of noise. Additionally, a comparison with a standard entanglement detection procedure will be presented to provide an intuition about the power of the few-copy approach in experimental settings and its applicability to large-scale quantum systems. Ultimately, we will show that not only does this method solve the problem of high demand in practice, but it also allows, in certain cases, to extract the fidelity of the quantum state --- i.e. to quantify the overlap between the experimental state and an ideal target state --- with no additional overhead. In this way, we will provide a perspective of how the few-copy method can prove particularly useful for the next generation of quantum devices.

\section{Detecting entanglement with few copies only}
\begin{figure}
\centering
\includegraphics[width=0.48\textwidth]{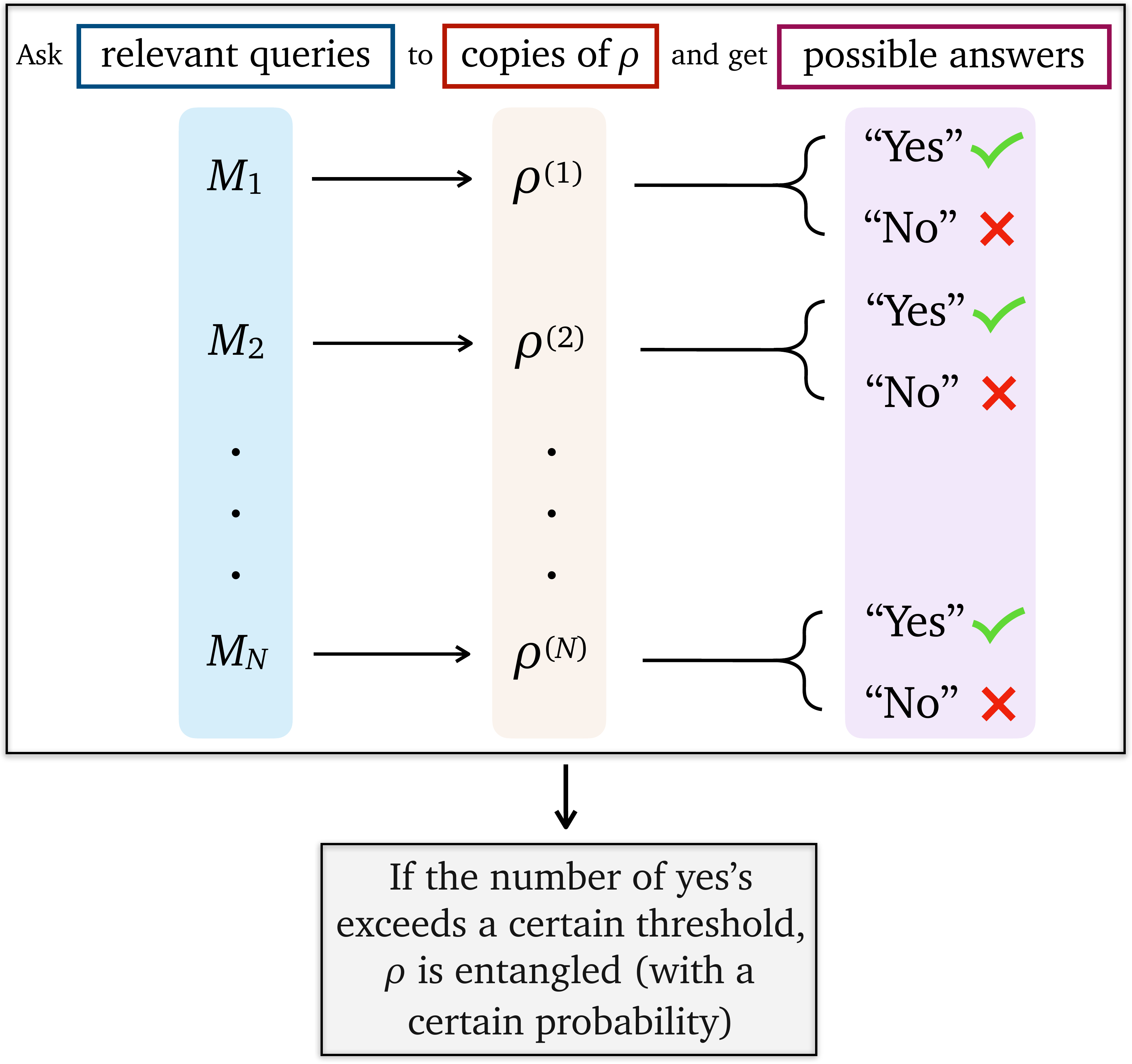}
\caption{\textbf{Theoretical protocol for detecting entanglement on a quantum state $\boldsymbol{\rho}$ in a probabilistic way.} $N$ copies of $\rho$ are queried with appropriate questions (corresponding to specific randomly sampled observables $M_i$) and produce yes/no answers (corresponding to 1/0 outcomes, respectively). Eventually, the number of positive answers is counted. If this number exceeds a certain threshold, the quantum state $\rho$ exhibits entanglement with a certain probability.}
\label{fig1}
\end{figure}

A recent probabilistic scheme has demonstrated that reliable entanglement detection is possible using a very limited amount of copies of a quantum state~\cite{saggio2019experimental}. The key idea behind this resource-saving approach is to formulate the task of entanglement detection on a certain quantum state $\rho$ in terms of a probabilistic procedure where entanglement is seen as the ability of $\rho$ to answer specific yes/no questions. The more ``yes'' answers are given, the higher the probability that $\rho$ contains entanglement. This idea is illustrated schematically in Figure~\ref{fig1}. The questions, indicated by $M_i$ with $i=1,...,L$, correspond to adequate observables that are drawn randomly from a set $\mathpzc{M}$ a certain number of times $N$ and are applied to copies of $\rho$. Their application produces a binary outcome $m_i$ that can take the value of either $1$ (``yes'') or $0$ (``no'') for each copy. If the number of positive answers exceeds a certain threshold, the quantum state $\rho$ exhibits entanglement with high probability. Both this threshold and probability will be quantified in the course of this section. 

The first step to apply the procedure is to construct the set $\mathpzc{M}$ of operators $M_i$ such that the probability to obtain positive answers (that is, outcomes $m_i=1$) for all separable states $\rho_{\mathrm{s}}$ is upper bounded by a certain value $p_{{\mathrm{s}}}<1$, also called the \emph{separable bound}. On the other hand, the same probability for a certain entangled state $\rho_{{\mathrm{e}}}$ (the target state) is maximized to a certain value $p_{{\mathrm{e}}}$ that is strictly greater than the separable bound $p_{{\mathrm{s}}}$. The value $p_{{\mathrm{e}}}$ is referred to as the \emph{entanglement value}. Assuming we can experimentally prepare a certain quantum state $\rho_{{\mathrm{exp}}}$, and that the application of $N$ randomly drawn operators $M_i$ to it returns $S$ outcomes $1$, we can therefore write the observed entanglement value as $p_{{\mathrm{e}}}^{\mathrm{obs}}=S/N$ and thus find its deviation from the separable bound $p_{\mathrm{s}}$ as
\begin{equation}
\delta=p_{{\mathrm{e}}}^{\mathrm{obs}} - p_{\mathrm{s}}=\frac{S}{N} - p_{\mathrm{s}}.
\label{deltaexp}
\end{equation}
If $S/N$ exceeds $p_{\mathrm{s}}$ (or equivalently if $S$ exceeds $N p_{\mathrm{s}}$, with the latter being the threshold mentioned in Figure~\ref{fig1}), entanglement is detected in $\rho_{\mathrm{exp}}$. Whenever this is the case, we associate ``success'' to the protocol. The last thing to figure out is the probability that this occurs (that is, the probability that $\delta$ evaluates to a positive number). One can see from Ref.~\cite{dimic2018single} that this probability of success, indicated by $P(\delta)$, is upper bounded as
\begin{equation}
P(\delta) \leq e^{-D(p_{\mathrm{s}}+\delta||p_{\mathrm{s}}) N},
\label{p_success_general}
\end{equation}
for any separable state $\rho_{\mathrm{s}}$. Here $D(x||y)$ is the Kullback-Leibler divergence and reads $D(x||y)=x \text{log} \frac{x}{y}+(1-x)\text{log} \frac{1-x}{1-y}$. Hence the probability or confidence $C(\delta)$ of detecting entanglement is lower bounded as 
\begin{equation}
C(\delta) = 1- P(\delta) \geq 1- e^{-D(p_{\mathrm{s}}+\delta||p_{\mathrm{s}}) N}=C_{\mathrm{min}}(\delta),
\label{confidence}
\end{equation}
which goes exponentially fast to unity in the number of outcomes $N$. It is important to highlight that, since each $M_i$ is applied to a copy of the quantum state, $N$ is also the number of interrogated state copies, or equivalently the number of experimental repetitions. It follows from Equation~(\ref{confidence}) that, for a fixed minimum confidence $C_{\mathrm{min}} \equiv C_0$, the number of copies required is estimated to be at most 
\begin{equation}
N_{\mathrm{max}}= \frac{-\mathrm{log}(1- C_0)}{D(p_{\mathrm{s}}+\delta||p_{\mathrm{s}})}.
\label{maxcopies}
\end{equation}
This means that this probabilistic framework allows one to detect entanglement in a quantum state $\rho$ using a maximum number of copies of $\rho$ that grows logarithmically at the rate of $D(p_{\mathrm{s}}+\delta||p_{\mathrm{s}})^{-1}$ as the confidence $C_0$ increases. In what follows we will show that $N_{\mathrm{max}}$ is a low number in general, hence the name \emph{few-copy} entanglement detection. This theoretical scheme was validated experimentally on a six-qubit photonic cluster state~\cite{saggio2019experimental}. It was shown that entanglement detection is possible with high confidence (more than $99\%$) using only about a hundred copies of the state.

Importantly, unlike standard approaches, this few-copy method also bypasses the need for independent and identically distributed (i.i.d.) copies of the quantum state through to the use of random sampling of measurement operators~\cite{dimic2018single,morris2021quantum}. As the generation of identical copies in each single experimental run is always challenging from a practical perspective, the fact that the i.i.d. assumption is not required  constitutes a clear advantage. 

Besides this few-copy case, which is the main focus of this article, it is worth mentioning another scenario (studied in detail in Ref.~\cite{dimic2018single}) embedded in this probabilistic framework, namely the case where the probability of success is upper bounded by a certain function $e^{-\alpha n}$, where $\alpha$ is a positive constant and $n$ is the number of qubits constituting the quantum state. As a consequence, when target states are prepared, the confidence for entanglement detection is at least $C_{\mathrm{min}}=1-e^{-\alpha n}$, which approaches 1 exponentially fast in $n$. This suggests that detecting entanglement with high confidence is possible even with just one copy of the state, provided that $n$ is sufficiently large. For example, it was shown that entanglement detection is possible with 95\% confidence using only one copy of a linear cluster state composed of $n=24$ qubits. It was proved that also k-producible states and ground states of local Hamiltonians are amenable to single-copy entanglement verification.

\subsection{Entanglement witnesses and the few-copy protocol}
\label{ewatfp}
The application of the few-copy protocol to an experimental state $\rho_{\mathrm{exp}}$ requires exceptionally little effort. The procedure can be summarised as follows:
\begin{itemize}
\item[\textit{(i)}] Randomly draw the operators $M_i$ (each with a given probability $\varepsilon_i$) from the set $\mathpzc{M}$ $N$ times; 
\item[\textit{(ii)}] Apply each of them to a copy of $\rho_{\mathrm{exp}}$ and note down the corresponding outcome $m_i$ ($0$ or $1$);
\item[\textit{(iii)}] Extract $S$ (the number of outcomes $m_i=1$) and use it to calculate the deviation $\delta$ in Equation~(\ref{deltaexp}). Note that the value $p_{\mathrm{s}}$ is obtained from the theoretical derivation and is therefore fixed and independent of the experimental implementation;
\item[\textit{(iv)}] If $\delta>0$, use it to evaluate the minimum confidence $C_{\mathrm{min}}(\delta)$ in Equation~(\ref{confidence}). This is the confidence with which entanglement is detected in $\rho_{\mathrm{exp}}$. If $\delta\leq0$, the protocol is inconclusive and therefore no information can be gained.
\end{itemize}
The only question left is how to initially derive the set $\mathpzc{M}$ and the separable bound $p_{\mathrm{s}}$. A general and formal method to extract these quantities is given in Refs.~\cite{saggio2019experimental,morris2021quantum}, which also contain many details and examples. In short, all is needed to apply this procedure is the definition of any so-called \textit{entanglement witness} operator $W$ tailored to the quantum state one wishes to probe. Once $W$ is defined, the so-called ``witness translation method'' outlined in Refs.~\cite{saggio2019experimental,morris2021quantum} can be used to derive $\mathpzc{M}$ and $p_{\mathrm{s}}$. Constructing such witnesses is nowadays a pretty standard technique, and in general many witnesses have been already derived for many different quantum states~\cite{toth2005entanglement,toth2005detecting,chen2007multiqubit,guhne2009entanglement}. 

This witness translation method makes use of entanglement witness operators in a new way. The standard witness-based technique (which is among the most common approaches for entanglement detection) requires the estimation of the expectation value $\avg{W}_{\rho}=\mathrm{Tr}[W\rho]$, where $\rho$ is the quantum state to probe. If $\avg{W}_{\rho}$ evaluates to a negative value, then $\rho$ is entangled. More formally, an entanglement witness $W$ is normalized such that
\begin{equation}
    \begin{cases}
      \avg{W}_{\rho_{\mathrm{s}}}\geq0 & \text{for all separable states $\rho_{\mathrm{s}}$,}\\
      \avg{W}_{\rho_{\mathrm{e}}}<0 & \text{for at least one entangled state $\rho_{\mathrm{e}}$.}
    \end{cases}
    \label{witnesssep}
\end{equation}

Although it may sound simple, this technique presents some substantial drawbacks as the size of the quantum state increases. First of all, witnesses are not directly measurable quantities, in the sense that they are in general not accessible locally~\cite{terhal2002detecting}. They have to be decomposed in terms of local observables $W_i$ as $W=\sum_{i=1}^{Q} W_i$, hence requiring the estimation of $Q$ expectation values in independent experimental runs. Therefore, if the experimental count rate for a certain experimental state $\rho$ is particularly low, it might be challenging (or even impossible) to estimate each $\avg{W_i}_{\rho}$ with low statistical error. Clearly, the more local terms are present in the witness decomposition (in general, the number of local observables increases exponentially with the number of qubits~\cite{toth2005detecting,bourennane2004experimental}), the worse this situation gets. For this reason, efforts have been made to reduce the number of terms composing entanglement witnesses~\cite{toth2005entanglement,toth2005detecting} or even the number of measurements via the use of adaptive strategies~\cite{garcia2021learning}. Despite these improvements, it is clear that measuring expectation values is not a practical task when the experimental count rate is particularly low. In such case, a consistent number of copies would still be needed for each local term $\avg{W_i}_{\rho}$ to be measured with high accuracy.

We stress here that while the few-copy approach does need to define an entanglement witness to build the theoretical protocol, its use is based on a completely different theoretical groundwork that provides a feasible tool to adequately claim the presence of entanglement in a quantum state bypassing measurements of expectation values. Essentially, the few-copy method detects entanglement through the use of probabilities of success, and thus drastically reduces the experimental requirements (number of detection events needed).

\section{An experimental perspective on noise tolerance}
To give a more quantitative analysis of how efficient the few-copy approach is (also compared to the standard witness method), we investigate how the experimental requirements vary according to the amount of noise affecting the quantum state. Such considerations come in light of the fact that experimental quantum states are always mixed with some noise that might have potential detrimental effects on the performance of the desired quantum protocol. Therefore, a careful noise analysis is essential to gather insights on its range of applicability.

A common way to account for noise in a quantum state is to consider a target (ideal) state $\rho_{\mathrm{t}}=\ket{\psi}\bra{\psi}_{\mathrm{t}}$ mixed with the white noise $\rho_{\mathrm{white}}=\mathbb{1}/d^n$, where $d$ is the dimension of the local Hilbert space and $n$ the number of qubits constituting the quantum state. This mixture is written as
\begin{equation}
\rho = \lambda \frac{\mathbb{1}}{d^n} + (1-\lambda) \ket{\psi}\bra{\psi}_{\mathrm{t}},
\label{statenoise}
\end{equation}
with $0<\lambda<1$ being the amount of noise in the state. In the absence of noise ($\lambda=0$), $\rho$ reduces to the target state $\ket{\psi}\bra{\psi}_{\mathrm{t}}$, while in the presence of maximal noise ($\lambda=1$), $\rho$ coincides with a maximally mixed state. Obviously, white noise is not the only way to model noise and the generated quantum state may not always be of the form of Equation~(\ref{statenoise}). However, this model is often a good approximation and offers a feasible tool to draw general conclusions about how the requirements for entanglement detection may vary in a reasonably realistic scenario. 

To give an intuition of how noise affects the few-copy protocol, we consider the example of a generic $n$-qubit cluster state $\ket{C}$ and the witness 
\begin{equation}
W_{\ket{C}}=\frac{1}{2} \mathbb{1} - \ket{C}\bra{C},
\label{witness}
\end{equation}
which is standardly defined to detect genuine multipartite entanglement in cluster states (and more in general in graph states~\cite{toth2005detecting}). We can always use the fact that a cluster state $\ket{C}$ can be decomposed in terms of its so-called \emph{stabilizers} $S_i^{\ket{C}}$, with $i=1,...,2^n$ as
\begin{equation}
\ket{C}\bra{C}=\frac{1}{2^n} \sum_{i=1}^{2^n} S_i^{\ket{C}}.
\label{clusterdecomposed}
\end{equation}
The stabilizers $S_i^{\ket{C}}$ are written as tensor products of Identity and Pauli operators and have the property that~\cite{browne2016one}
\begin{equation}
S_i^{\ket{C}} \ket{C}=\ket{C}.
\label{stabilizer}
\end{equation}
At this point we can write the decomposition in Equation~(\ref{clusterdecomposed}), and therefore the witness $W_{\ket{C}}$ in Equation~(\ref{witness}), in terms of $2^n$ local binary observables $M_i=\frac{\mathbb{1}+S_i^{\ket{C}}}{2}$ (constituting the set $\mathpzc{M}$), and then make use of the first property in~(\ref{witnesssep}) to derive the separable bound $p_{\mathrm{s}}$. Following this procedure, we find $p_{\mathrm{s}}=3/4$. 
Note that while this example summarises only the case of a specific witness, a general method is given in Ref.~\cite{saggio2019experimental} to translate any witness into the probabilistic framework.

We now mix the cluster state $\ket{C}$ with the white noise according to Equation~(\ref{statenoise}), thus obtaining
\begin{equation}
\rho_{\ket{C}} = \lambda \frac{\mathbb{1}}{2^n} + (1-\lambda) \ket{C}\bra{C},
\label{clusternoise}
\end{equation}
where we used the fact that $d=2$. 

Let us start with considering the ideal case where $\lambda=0$. We can obtain the entanglement value $p_{\mathrm{e}}^{\mathrm{obs}}=S/N$ by following the steps \textit{(i)-(iv)} listed in Subsection~\ref{ewatfp}. As the $S_i^{\ket{C}}$ stabilize the state (i.e. Equation~(\ref{stabilizer}) holds), all $M_i=1$ deterministically and thus $p_{\mathrm{e}}^{\mathrm{obs}}=1$. Therefore, the deviation from the separable bound can be obtained from Equation~(\ref{deltaexp}) and reads $\delta=1/4$. We can now plug $\delta$ and $p_{\mathrm{s}}$ in Equation~(\ref{maxcopies}) to estimate the number of copies that would be needed at most to reach a fixed confidence $C_0$ of, for example, 0.99. We find $N_{\mathrm{max}}\approx 16$, which is an exceptionally low number. We stress here that this number is independent of the number of qubits, and will therefore be valid for cluster states, and more in general graph states, of any dimension.

While this certainly provides a practical intuition about the power of this few-copy approach, such ideal case can never be reached in practice. When noise is present in the system, the quantity $p_{\mathrm{e}}^{\mathrm{obs}}=S/N$ does no longer equal $1$ deterministically. Considering the noisy state in Equation~(\ref{clusternoise}), we now obtain 
\begin{equation}
p_{\mathrm{e}}^{\mathrm{obs}}(\lambda)=\lambda p_{\frac{\mathbb{1}}{2^n}} + (1-\lambda) p_{\ket{C} \bra{C}},
\label{frequencynoise}
\end{equation}
where $p_{\frac{\mathbb{1}}{2^n}}$ and $p_{\ket{C} \bra{C}}$ are the probabilities of obtaining outcome $m_i=1$ from the random sampling and application of the observables $M_i$ for the case of the totally mixed state $\mathbb{1}/{2^n}$ and the ideal target state $\ket{C} \bra{C}$, respectively. Given that the specific decomposition of the target state includes $2^n$ operators of which one is the Identity $\mathbb{1}^{\otimes n}$ (which deterministically gives outcome 1), we obtain that
\begin{equation}
p_{\frac{\mathbb{1}}{2^n}}=\frac{1+\frac{1}{2}(2^n-1)}{2^n},
    \label{bound}
\end{equation}
because all the other $2^n-1$ operators return outcome $1$ with probability $1/2$ each when applied to the totally mixed state. The probability $p_{\ket{C} \bra{C}}$ is instead $1$ deterministically for states of any dimension. Therefore, substituting $p_{\frac{\mathbb{1}}{2^n}}$ from Equation~(\ref{bound}) in Equation~(\ref{frequencynoise}), and considering $p_{\ket{C}\bra{C}}=1$, we find that 
\begin{equation}
p_{\mathrm{e}}^{\mathrm{obs}}(\lambda) = 1+\lambda \frac{1-2^n}{2 \cdot 2^n},
\label{pnoisen}
\end{equation}
which becomes
\begin{equation}
p_{\mathrm{e}}^{\mathrm{obs}}(\lambda) \approx 1-\frac{\lambda}{2}
\label{pnoiselim}
\end{equation}
when $n$ is sufficiently large. 
Therefore, we find that the observed entanglement value scales linearly with the noise $\lambda$. Since the few-copy protocol is successful only for positive deviations $\delta$, we impose $\delta(\lambda)=p_{\mathrm{e}}^{\mathrm{obs}}(\lambda) -p_{\mathrm{s}}>0$ and find that the noise has to obey the condition
\begin{equation}
\lambda< \frac{2^n}{2(2^n-1)}=\lambda_{\mathrm{lim}}.
\label{lambdalim}
\end{equation}
Therefore, the maximum amount of noise tolerated by the protocol has be smaller than $\lambda_{\mathrm{lim}} \approx 1/2$ for large $n$.

\begin{figure}
\centering
\includegraphics[width=0.47\textwidth]{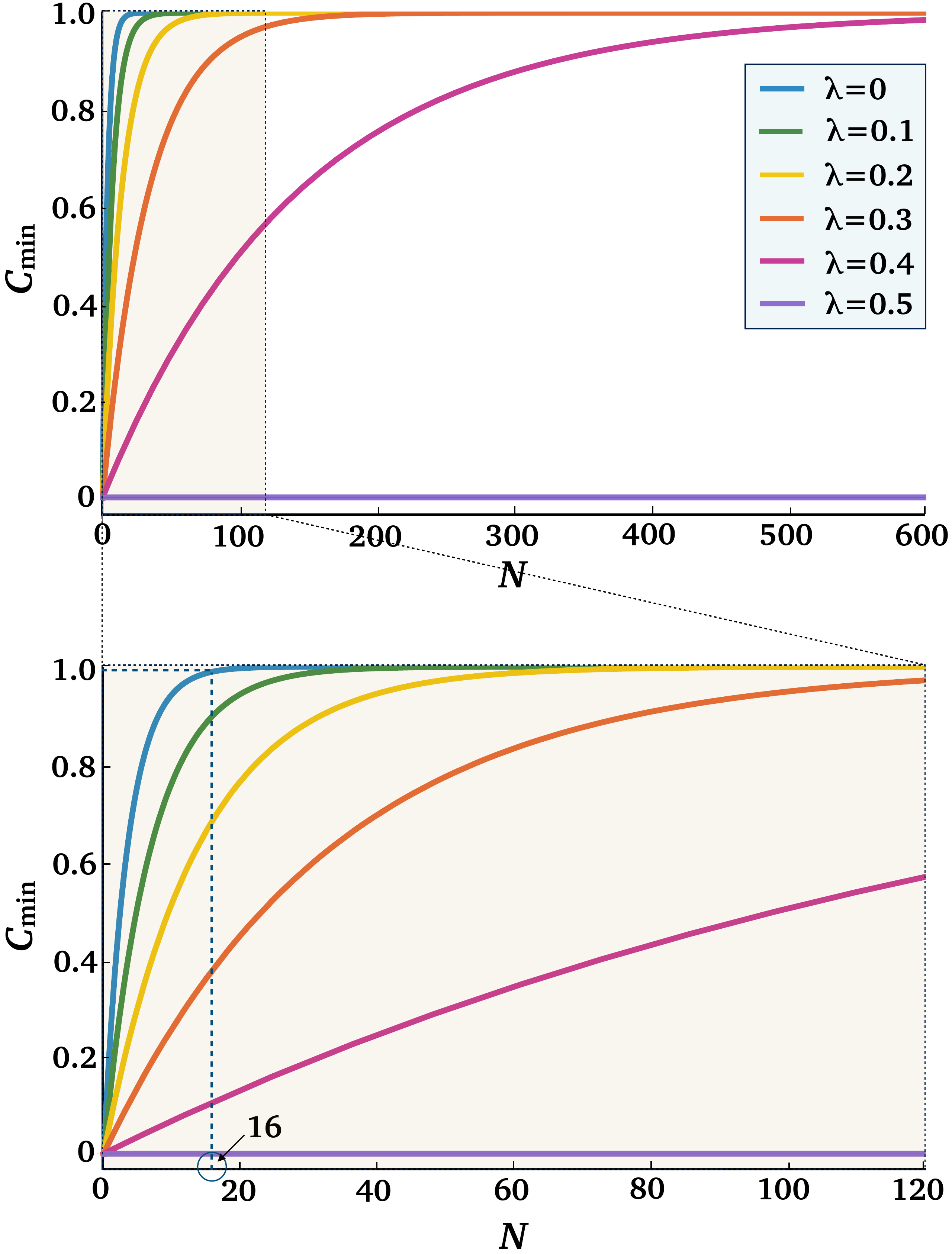}
\caption{\textbf{Minimum confidence growth versus number of copies for different noise levels $\boldsymbol{\lambda}$.} In the absence of noise ($\lambda=0$), the minimum confidence grows as $C_{\mathrm{min}}=1-p_{\mathrm{s}}^N$, where $p_{\mathrm{s}}=3/4$ in our example. In this case, as better visible from the zoomed plot, only $16$ copies are required to achieve a minimum confidence of about $0.99$. Increasing the amount of noise causes an increase in the number of copies necessary to achieve the same confidence according to Equation~(\ref{maxcopiesnoise}).}
\label{fig2}
\end{figure}

Once the scaling in Equation~(\ref{pnoiselim}) is obtained, we can use this result to study how the growth of the minimum confidence is affected by $\lambda$. Using the fact that $p_{\mathrm{s}}=3/4$, we find from Equation~(\ref{confidence}) that
\begin{equation}
C_{\mathrm{min}}(\lambda) \approx 1- e^{-D(1-\frac{\lambda}{2}||\frac{3}{4}) N}.
\label{confidencelambda}
\end{equation}
This growth is shown in Figure~\ref{fig2} for different values of $\lambda$.
As better visible from the zoomed plot, only $16$ copies are needed at most to achieve a confidence of about $0.99$ in the noiseless case. As the amount of noise increases, and for a fixed confidence $C_0$, we derive from Equation~(\ref{confidencelambda}) that the maximum number of copies scales as 
\begin{equation}
N_{\mathrm{max}}(\lambda)= \frac{-\mathrm{log}(1- C_0)}{D(1-\frac{\lambda}{2}||p_{\mathrm{s}})}.
\label{maxcopiesnoise}
\end{equation}
Therefore, even in the presence of a significant amount of noise, detecting genuine multipartite entanglement with high confidence is possible with about a few hundred copies on average. The limiting case of $\lambda=0.5$ shows a confidence fixed at zero. This is because at $\lambda=0.5$ the entanglement value equals the separable bound on average, and thus $\delta$ goes to zero. This prevents the confidence from growing.

\begin{figure*}
\centering
\includegraphics[width=0.99\textwidth]{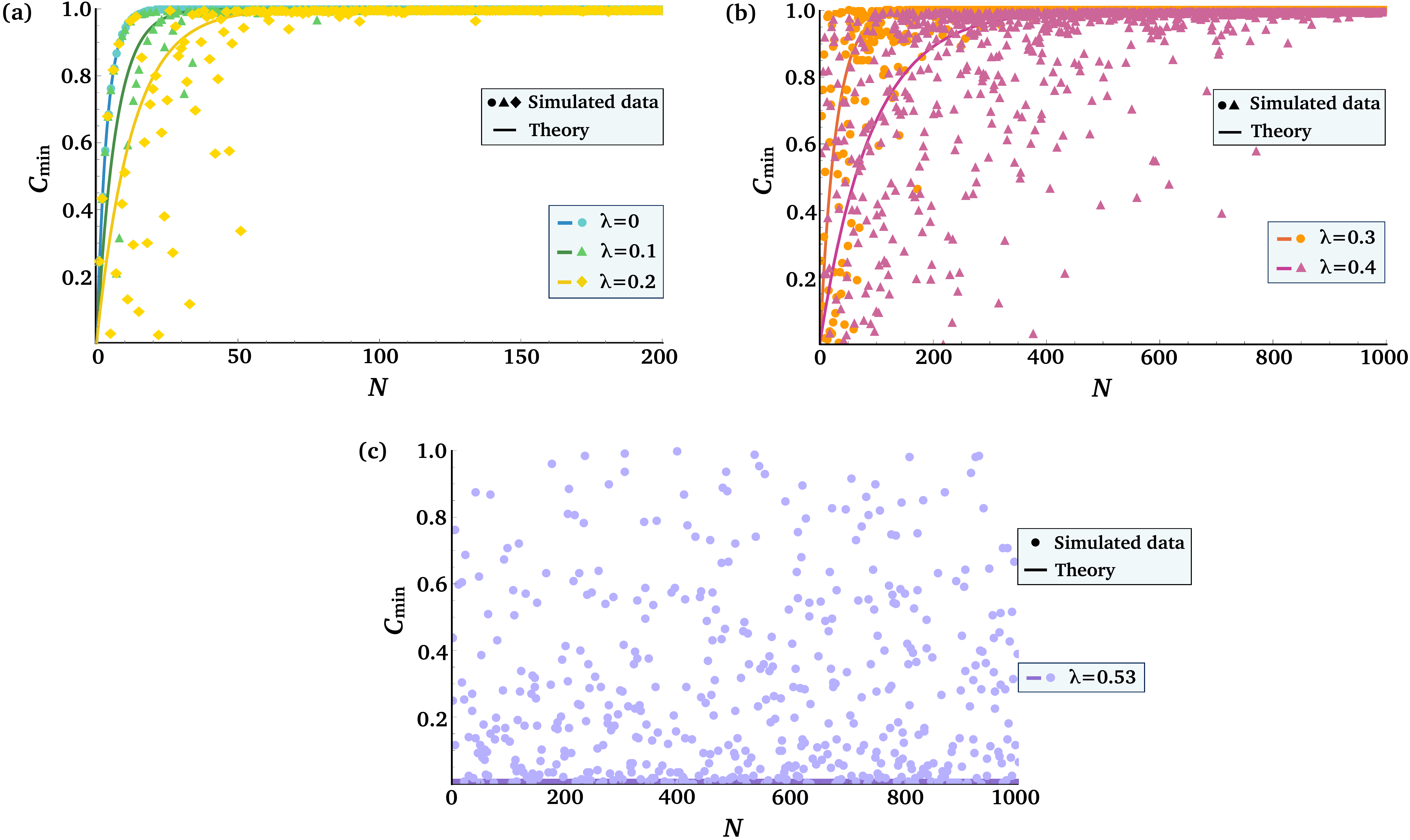}
\caption{\textbf{Minimum confidence growth versus number of copies for different noise levels $\boldsymbol{\lambda}$ in the simulated case of a $\boldsymbol{4}$-qubit linear cluster state.} The dots represent the simulated data, while the solid lines reproduce the theory. (a) and (b) show how the minimum confidence converges to unity with the number of copies when varying $\lambda$ from $0$ to $0.4$. The limiting case shown in (c) for $\lambda=0.53$ features no growth in the confidence (the data points present the same distribution for each value of $N$). Only the points for which $\delta>0$ have been used for the calculation of the confidence.}
\label{fig3}
\end{figure*}

To observe more in detail how the few-copy protocol works in a realistic scenario, we report a simulation of how the confidence is built up in an experimental setting. We consider the example of the following four-qubit state:
\begin{equation}
\ket{C_4}=\frac{1}{2} (\ket{0000}+\ket{0011}+\ket{1100}-\ket{1111}),
\label{cluster}
\end{equation}
which is equivalent to a four-qubit linear cluster state up to Hadamard transformations on the first and fourth qubit. After decomposing the corresponding witness --- that is, $W_{\ket{C_4}}=\frac{1}{2} \mathbb{1} - \ket{C_4}\bra{C_4}$ from Equation~(\ref{witness}) --- into $2^4=16$ binary local observables, we can simulate the entanglement detection procedure and find the plots reported in Figure~\ref{fig3}, where the dots correspond to the simulated data and the solid lines reproduce the theoretical predictions (true values of $C_{\mathrm{min}}$). Every time an outcome $0$ is obtained, the confidence is pulled down, and it is pulled up again when outcomes $1$ are returned. Obviously, increasing the number of copies suppresses these fluctuations, as visible in Figure~\ref{fig3}(a) and (b) (note that in the noiseless case the data points never deviate from the theoretical curve, because no outcomes $0$ are ever obtained). Figure~\ref{fig3}(a) shows that about $200$ copies are required at most to reach high confidence when $\lambda<0.2$. As $\lambda$ increases, more and more copies are needed, as shown in Figure~\ref{fig3}(b). At $\lambda=0.53$, which is the theoretical noise limit calculated from Equation~(\ref{lambdalim}) for $n=4$, it is visible from Figure~\ref{fig3}(c) that no convergence takes place. In this case, the data points are equally distributed for any value of $N$.

Let us now study the case where information about the presence of entanglement is inferred from measuring the expectation value of the witness in Equation~(\ref{witness}). Also in this case, a maximum amount of tolerated noise can be derived. This is achieved by expanding the witness $\avg{W_{\ket{C}}}=\text{Tr}  [W_{\ket{C}} \rho ]$ using Equation~(\ref{statenoise}) with $d=2$ and $\ket{\psi}_{\mathrm{t}}=\ket{C}$. Using the first property in~(\ref{witnesssep}), it is easy to see that the following condition must hold~\cite{toth2005entanglement}:
\begin{equation}
\lambda < \lambda_{\mathrm{lim}}= \frac{-\text{Tr}[W_{\ket{C}} \ket{C} \bra{C}]}{\frac{\text{Tr}[ W_{\ket{C}}]}{2^n}-\text{Tr}[W_{\ket{C}} \ket{C} \bra{C}]} = \frac{2^n}{2 (2^n-1)}
    \label{noisemaxwitness}
\end{equation}
which equals Inequality~(\ref{lambdalim}). This means that in such case both the few-copy protocol and the standard witness approach feature the same tolerance to noise.

However, contrarily to the few-copy method, we now have to set the accuracy we wish to reach in our measurement, as we are performing parameter estimation. As already mentioned in the previous section, entanglement witnesses are generally decomposed into local terms. This means that we need to estimate, with a certain degree of accuracy, as many expectation values as there are in the witness decomposition. As an example, let us assume that we wish to estimate the expectation value of an observable $X$ (it could be, for example, one of the stabilizers $S_i^{\ket{C_4}}$) with an accuracy of $\varepsilon=0.02$ and a confidence level of $95\%$. It is obvious enough that the desired accuracy and confidence level strongly depend on the specific experimental settings and may thus differ. Our aim here is to provide a basic intuition of how the experimental requirements might vary from one protocol to another through the study of general examples. Carrying out a simulation of the measurement process of $\avg{X}$ with and without noise, we find the number of experimental repetitions needed to achieve accuracy of $\varepsilon=0.02$ and confidence level of $95\%$ to vary from a minimum of about a thousand for $\lambda=0.1$, to a maximum of a few thousands for $\lambda=0.5$. The accuracy is derived from propagated Poissonian statistics of the detection events. To understand what this may imply experimentally, let us provide an example by referring to the experimental conditions in Ref.~\cite{zhong201812}, where the entanglement of a 12-photon state was proven. While surely technologically impressive, the setup presented by the authors has the capability of generating about one copy of the quantum state per hour. Clearly, collecting thousands of copies to achieve the aforementioned precision and confidence level would be totally out of reach. Fortunately, such high requirements are not always needed, and often estimation of expectation values with a confidence level of $68\%$ is good enough to prove the presence of entanglement, as shown for example in Refs.~\cite{zhong201812,wang2016experimental,lu2007experimental}. However, still in the order of a few hundred to a few thousand detection events would be required in this case per local observable (depending on the desired accuracy $\varepsilon$). If the entanglement witness is decomposed into $Q$ local terms, then one would need to repeat this procedure $Q$ times, using $Q$ times more resources. While it is true that parameter estimation reveals more information about a quantum state, its experimental implementation has to account for a proper statistical analysis through the collection of a sufficient number of detection events. Instead, the few-copy protocol is based on a probabilistic assumption, and the scaling in the number of resources is independent of the accuracy $\varepsilon$ (because no parameter estimation is performed). 

\section{Fidelity estimation}
In addition to the detection of quantum entanglement, we highlight that in some cases the few-copy protocol also provides a direct way to estimate the fidelity $F$ between the experimental (noisy) state $\rho_{\mathrm{exp}}$ and the target state  $\rho_{\mathrm{t}}$. When treating pure states $\rho_{\mathrm{exp}}= \ket{\psi_{\mathrm{exp}}} \bra{\psi_{\mathrm{exp}}}$ and $\rho_{\mathrm{t}}= \ket{\psi_{\mathrm{t}}}\bra{\psi_{\mathrm{t}}}$, $F$ can be quantified as $F= \langle \psi_{\mathrm{t}} | \rho_{\mathrm{exp}}| \psi_{\mathrm{t}} \rangle = |\bra{\psi_{\mathrm{exp}}} \psi_{\mathrm{t}}\rangle |^2$. Such possibility is definitely a useful feature, given that in many cases it is essential to quantify how close an experimental state is to its target. As an example, given an $n$-qubit cluster state $\ket{C}$ (the target state), fidelity estimation can be performed with the generic witness $W_{\ket{C}}$ defined in Equation~(\ref{witness}) employing exactly the same resources needed to estimate entanglement in the same state. This procedure is essentially related to the direct fidelity estimation protocol~\cite{flammia2011direct}. After decomposing the cluster state in terms of its stabilizers and then rewriting it in terms of the local binary observables $M_i$, one finds that~\cite{saggio2019experimental}
\begin{equation}
| C \rangle \langle C | = \sum_{i=1}^{2^n} \frac{1}{2^n} (2 M_i -\mathbb{1}).
\label{decompositionfidelity}
\end{equation}
From here, calculating its expectation value directly leads to the fidelity 
\begin{equation}
F=\langle C | \rho_{\mathrm{exp}} | C\rangle = 2\frac{\sum_{i=1}^{2^n}\avg{M_i}}{2^n}-1.
\label{fidelity}
\end{equation}
Therefore, once the entanglement detection protocol has been implemented, the same data can be used to estimate the fidelity as well. In this sense, both tasks come at the same cost, hence requiring no additional overhead in the experimental implementation.

\section{Concluding remarks}
In this article, we have shown a few-copy probabilistic method that can detect entanglement in arbitrarily large quantum systems in an efficient and reliable way even in the presence of noise. The main theoretical requirement needed to apply this method is simply the definition of an entanglement witness for the quantum state under consideration. Once a witness is constructed (this task is nowadays pretty well defined for many quantum states), the derivation of the binary observables and of the separable bound is general and straightforward. We stress here that any quantum state for which a witness is defined is well-suited for efficient probabilistic entanglement detection. The experimental requirements reduce instead to the application of few local measurements to each qubit constituting the quantum state. As such, whenever individual addressability of each qubit is possible, probabilistic entanglement detection appears to be a feasible task even for large and noisy quantum systems. 

We have also shown how the presence of noise affects the protocol. Although both the few-copy method and the standard witness-based technique are limited by a certain noise threshold, the former requires significantly less copies than the latter, and it is therefore more suited for experimental settings that, for example, suffer from a low count rate or do not produce identically and independent copies.

Moreover, it is shown in Ref.~\cite{saggio2019experimental} that this method can distinguish between genuine multipartite entanglement and only some entanglement in the quantum system (intuitively enough, the requirements for the detection of only some entanglement will be even lower compared to genuine multipartite entanglement detection). This feature is clearly essential for the implementation of quantum computation protocols requiring specific entangled quantum states.

Therefore, as the race for the next generation of NISQ devices has already begun, the few-copy approach may truly have the potential to mitigate the experimental costs while still being robust against the presence of noise. 

\medskip

\medskip
\textbf{Acknowledgements} \par 
The authors are thankful to Borivoje Daki\'c for helpful discussions. V.S. and P.W. acknowledge support from the Austrian Science Fund (FWF) through BeyondC (F7113) and Research Group (FG 5), the research platform TURIS, the European Commission through EPIQUS (no. 899368), and AppQInfo (no. 956071).

\bibliographystyle{unsrturl}
\bibliography{biblio_MSP}

\begin{thebibliography}{10}

\bibitem{einstein1935can}
Albert Einstein, Boris Podolsky, and Nathan Rosen.
\newblock Can quantum-mechanical description of physical reality be considered
  complete?
\newblock {\em Physical review}, 47(10):777, 1935.

\bibitem{bell1964einstein}
John~S Bell.
\newblock On the einstein podolsky rosen paradox.
\newblock {\em Physics Physique Fizika}, 1(3):195, 1964.

\bibitem{aspect1981experimental}
Alain Aspect, Philippe Grangier, and G{\'e}rard Roger.
\newblock Experimental tests of realistic local theories via bell's theorem.
\newblock {\em Physical review letters}, 47(7):460, 1981.

\bibitem{aspect1982experimental}
Alain Aspect, Philippe Grangier, and G{\'e}rard Roger.
\newblock Experimental realization of einstein-podolsky-rosen-bohm
  gedankenexperiment: a new violation of bell's inequalities.
\newblock {\em Physical review letters}, 49(2):91, 1982.

\bibitem{aspect1982experimental_}
Alain Aspect, Jean Dalibard, and G{\'e}rard Roger.
\newblock Experimental test of bell's inequalities using time-varying
  analyzers.
\newblock {\em Physical review letters}, 49(25):1804, 1982.

\bibitem{jozsa2003role}
Richard Jozsa and Noah Linden.
\newblock On the role of entanglement in quantum-computational speed-up.
\newblock {\em Proceedings of the Royal Society of London. Series A:
  Mathematical, Physical and Engineering Sciences}, 459(2036):2011--2032, 2003.

\bibitem{morales2015quantum}
Guillermo Morales-Luna.
\newblock Quantum communication protocols based on entanglement swapping.
\newblock In {\em Journal of Physics: Conference Series}, volume 624, page
  012003. IOP Publishing, 2015.

\bibitem{bennett1993teleporting}
Charles~H Bennett, Gilles Brassard, Claude Cr{\'e}peau, Richard Jozsa, Asher
  Peres, and William~K Wootters.
\newblock Teleporting an unknown quantum state via dual classical and
  einstein-podolsky-rosen channels.
\newblock {\em Physical review letters}, 70(13):1895, 1993.

\bibitem{bouwmeester1997experimental}
Dik Bouwmeester, Jian-Wei Pan, Klaus Mattle, Manfred Eibl, Harald Weinfurter,
  and Anton Zeilinger.
\newblock Experimental quantum teleportation.
\newblock {\em Nature}, 390(6660):575--579, 1997.

\bibitem{pan1998experimental}
Jian-Wei Pan, Dik Bouwmeester, Harald Weinfurter, and Anton Zeilinger.
\newblock Experimental entanglement swapping: entangling photons that never
  interacted.
\newblock {\em Physical review letters}, 80(18):3891, 1998.

\bibitem{briegel2009measurement}
Hans~J Briegel, David~E Browne, Wolfgang D{\"u}r, Robert Raussendorf, and
  Maarten Van~den Nest.
\newblock Measurement-based quantum computation.
\newblock {\em Nature Physics}, 5(1):19--26, 2009.

\bibitem{walther2005experimental}
Philip Walther, Kevin~J Resch, Terry Rudolph, Emmanuel Schenck, Harald
  Weinfurter, Vlatko Vedral, Markus Aspelmeyer, and Anton Zeilinger.
\newblock Experimental one-way quantum computing.
\newblock {\em Nature}, 434(7030):169--176, 2005.

\bibitem{friis2019entanglement}
Nicolai Friis, Giuseppe Vitagliano, Mehul Malik, and Marcus Huber.
\newblock Entanglement certification from theory to experiment.
\newblock {\em Nature Reviews Physics}, 1(1):72--87, 2019.

\bibitem{preskill2018quantum}
John Preskill.
\newblock Quantum computing in the nisq era and beyond.
\newblock {\em Quantum}, 2:79, 2018.

\bibitem{james2005measurement}
Daniel~FV James, Paul~G Kwiat, William~J Munro, and Andrew~G White.
\newblock On the measurement of qubits.
\newblock In {\em Asymptotic Theory of Quantum Statistical Inference: Selected
  Papers}, pages 509--538. World Scientific, 2005.

\bibitem{toth2010permutationally}
G{\'e}za T{\'o}th, Witlef Wieczorek, David Gross, Roland Krischek, Christian
  Schwemmer, and Harald Weinfurter.
\newblock Permutationally invariant quantum tomography.
\newblock {\em Physical review letters}, 105(25):250403, 2010.

\bibitem{gross2010quantum}
David Gross, Yi-Kai Liu, Steven~T Flammia, Stephen Becker, and Jens Eisert.
\newblock Quantum state tomography via compressed sensing.
\newblock {\em Physical review letters}, 105(15):150401, 2010.

\bibitem{morris2019selective}
Joshua Morris and Borivoje Daki{\'c}.
\newblock Selective quantum state tomography.
\newblock {\em arXiv preprint arXiv:1909.05880}, 2019.

\bibitem{aaronson2019shadow}
Scott Aaronson.
\newblock Shadow tomography of quantum states.
\newblock {\em SIAM Journal on Computing}, 49(5):STOC18--368, 2019.

\bibitem{huang2020predicting}
Hsin-Yuan Huang, Richard Kueng, and John Preskill.
\newblock Predicting many properties of a quantum system from very few
  measurements.
\newblock {\em Nature Physics}, 16(10):1050--1057, 2020.

\bibitem{flammia2011direct}
Steven~T Flammia and Yi-Kai Liu.
\newblock Direct fidelity estimation from few pauli measurements.
\newblock {\em Physical review letters}, 106(23):230501, 2011.

\bibitem{guhne2009entanglement}
Otfried G{\"u}hne and G{\'e}za T{\'o}th.
\newblock Entanglement detection.
\newblock {\em Physics Reports}, 474(1-6):1--75, 2009.

\bibitem{saggio2019experimental}
Valeria Saggio, Aleksandra Dimi{\'c}, Chiara Greganti, Lee~A Rozema, Philip
  Walther, and Borivoje Daki{\'c}.
\newblock Experimental few-copy multipartite entanglement detection.
\newblock {\em Nature physics}, 15(9):935--940, 2019.

\bibitem{dimic2018single}
Aleksandra Dimi{\'c} and Borivoje Daki{\'c}.
\newblock Single-copy entanglement detection.
\newblock {\em npj Quantum Information}, 4(1):1--8, 2018.

\bibitem{morris2021quantum}
Joshua Morris, Valeria Saggio, Aleksandra Go{\v{c}}anin, and Borivoje
  Daki{\'c}.
\newblock Quantum verification with few copies.
\newblock {\em arXiv preprint arXiv:2109.03860}, 2021.

\bibitem{toth2005entanglement}
G{\'e}za T{\'o}th and Otfried G{\"u}hne.
\newblock Entanglement detection in the stabilizer formalism.
\newblock {\em Physical Review A}, 72(2):022340, 2005.

\bibitem{toth2005detecting}
G{\'e}za T{\'o}th and Otfried G{\"u}hne.
\newblock Detecting genuine multipartite entanglement with two local
  measurements.
\newblock {\em Physical review letters}, 94(6):060501, 2005.

\bibitem{chen2007multiqubit}
Lin Chen and Yi-Xin Chen.
\newblock Multiqubit entanglement witness.
\newblock {\em Physical Review A}, 76(2):022330, 2007.

\bibitem{terhal2002detecting}
Barbara~M Terhal.
\newblock Detecting quantum entanglement.
\newblock {\em Theoretical Computer Science}, 287(1):313--335, 2002.

\bibitem{bourennane2004experimental}
Mohamed Bourennane, Manfred Eibl, Christian Kurtsiefer, Sascha Gaertner, Harald
  Weinfurter, Otfried G{\"u}hne, Philipp Hyllus, Dagmar Bru{\ss}, Maciej
  Lewenstein, and Anna Sanpera.
\newblock Experimental detection of multipartite entanglement using witness
  operators.
\newblock {\em Physical review letters}, 92(8):087902, 2004.

\bibitem{garcia2021learning}
Guillermo Garc{\'\i}a-P{\'e}rez, Matteo~AC Rossi, Boris Sokolov, Francesco
  Tacchino, Panagiotis~Kl Barkoutsos, Guglielmo Mazzola, Ivano Tavernelli, and
  Sabrina Maniscalco.
\newblock Learning to measure: Adaptive informationally complete generalized
  measurements for quantum algorithms.
\newblock {\em Prx quantum}, 2(4):040342, 2021.

\bibitem{browne2016one}
Dan Browne and Hans Briegel.
\newblock One-way quantum computation.
\newblock {\em Quantum Information: From Foundations to Quantum Technology
  Applications}, pages 449--473, 2016.

\bibitem{zhong201812}
Han-Sen Zhong, Yuan Li, Wei Li, Li-Chao Peng, Zu-En Su, Yi~Hu, Yu-Ming He, Xing
  Ding, Weijun Zhang, Hao Li, et~al.
\newblock 12-photon entanglement and scalable scattershot boson sampling with
  optimal entangled-photon pairs from parametric down-conversion.
\newblock {\em Physical review letters}, 121(25):250505, 2018.

\bibitem{wang2016experimental}
Xi-Lin Wang, Luo-Kan Chen, Wei Li, H-L Huang, Chang Liu, Chao Chen, Y-H Luo,
  Z-E Su, Dian Wu, Z-D Li, et~al.
\newblock Experimental ten-photon entanglement.
\newblock {\em Physical review letters}, 117(21):210502, 2016.

\bibitem{lu2007experimental}
Chao-Yang Lu, Xiao-Qi Zhou, Otfried G{\"u}hne, Wei-Bo Gao, Jin Zhang,
  Zhen-Sheng Yuan, Alexander Goebel, Tao Yang, and Jian-Wei Pan.
\newblock Experimental entanglement of six photons in graph states.
\newblock {\em Nature physics}, 3(2):91--95, 2007.

\end{thebibliography}
\end{document}